# Interpretations of Rényi entropies and divergences

Peter Harremoës

*University of Copenhagen*

## Abstract

In this paper a new operationel definition of Rényi divergence and Rényi entropy is presented. Other operationel definitions are mentioned.

# I. INTRODUCTION

Shannon introduced his entropy in1948 [11] and the scientific community soon realized how powerful the new concept was. Shannon gave an axiomatic characterization of entropy and one should have thought that this would have settled the matter, but since then a number of entropy like quantities have appeared in the scientific literature. All these new quantities share some but not all properties with the Shannon entropy. The most important examples are the Rényi entropy and the Tsallis entropy. There are so many other definitions of entropy like quantities that Arndt [1] was able to write a whole book entitled "Information Measures". Although the book has more than 300 pages one may be surprised that Tsallis entropy is only mentioned shortly and only under the alternative name Havrda Chervat entropy. Obviously this short paper can only present a few results on this huge topic and give some pointers to the literature.

The focus in this paper will be interpretations via information theory. The fact is that most of the alternative definitions of entropy may be useful in very special situations or may fulfill modified axiom systems but they do not have operational definitions. Some of the alternative definitions will find applications and operational definitions in the future and most of them has already been more or less forgotten. This was what happened with the Havrda-Charvat entropy. It is defined in 1967 [6] and was forgotten and then rediscovered by C. Tsallis in 1988 [12]. Before we go any further it may be useful to specify what we mean by an operational definition. To us an operational definition of a quantity means that the quantity is the natural way to answer a natural question and that the quantity can be estimated by feasible measurements combined with combined with a reasonable number of computations. In this sense the Shannon entropy has a operational definition as a compression rate and Kolmogorov entropy has an operational definition as shortest program describing data.

The Rényi entropy was introduced by Rényi [9] and soon after it found application in graph theory. The original reason for Rényi to introduce his new entropy is said to be that he planned to use it in an information theoretic proof of the Central Limit Theorem. An information theoretic approach to the Central Limit Theorem dates back to Linnik [7], but his paper is very hard to understand. Rényi did not finish his work on the Central Limit Theorem before his all to early death, but his entropy has found a number of other applications.

In 1988 C. Tsallis reinvented the Havrda-Charvat entropy in order to generalize the well established Boltzmann-Gibb approach to statistical mechanics. A rich literature has developed since then and the original paper of Tsallis has more than 1000 references. The concept is mainly used in physics and has important applications for instance in thermodynamics of systems confined to surface of fractal dimension.

In information theory it has for long been known that Rényi entropies and divergences are related to so-called cutoff rates, see [3], [2] and [5]. Rényi entropies of order greater than 2 are also known to be related to search problems, see [10] and [8].

## II. CODING

Information is always *information about something*. The *description* of information must be distinguished from this "something", just as the words used to describe a dog are different from the dog itself. Description of information in precise technical terms is important since, in Shannon's words, cf. [11], it will allow *"reproducing at one point either exactly or approximately a message selected at another point"*. The descriptions in information theory are called *codes*.

An information *source* is some device or mechanism which generates elements from a certain set, the source *alphabet* $\mathbb{A}$ For simplicity we shall assume that $\mathbb{A}$ is finite. Table 1 shows a *code book* related to a source which generates a vowel of the English alphabet. The various *code-words* may be taken as a way to *represent*, indeed to *code*, the vowels. Or we may conceive the code-book as a strategy for obtaining information about the actual vowel via a series of yes/no questions. In our example, the first question will be "is the letter one of *a, o, u* or *y*?" . This corresponds to a "1" as the first *binary digit* – or *bit* as we shall say – in the actual code-word. Continuing asking questions related to the further bits, we end up by knowing the actual vowel. The number of bits required in order to identify a vowel is the *code-word length*, i.e. the number of bits in the corresponding code-word.

To ensure unambiguous identification, we require that a code is uniquely decodable. An important class of uniquely decodable codes is the set of *prefix-free codes*, i.e. no code-word in the code-book must be the beginning of another. Denoting code-word lengths by

| vowel | code-word | code-word length |
|-------|-----------|------------------|
| a | 11 | 2 |
| e | 00 | 2 |
| i | 01 | 2 |
| o | 100 | 3 |
| u | 1010 | 4 |
| y | 1011 | 4 |

TABLE I: Codebook for vowels in English.

$l_a$ ; $a \in \mathbb{A}$, we realize that *Kraft's inequality*

$$\sum_{a \in \mathbb{A}} 2^{-l_a} \leq 1 \qquad (1)$$

must hold – indeed, the binary subintervals of the unit interval which correspond, via successive bisections, to the various code-words must be pairwise disjoint, hence have total length at most 1. And, in the other direction, if numbers $l_a$ are given satisfying (1) then there exists a prefix-free code with the prescribed $l_x$'s as code-word lengths. By a more involved argument one can prove that Kraft's Inequality must hold for all uniquely decodable codes [4, Thm. 5.5.1].

Coding is partly of a combinatorial nature due to the requirement of integers as code-word lengths. For theoretical discussions it is desirable to "take the combinatorial dimension out of coding". This can be done by allowing arbitrary real numbers as code-word lengths. We therefore define an *idealized code over the alphabet* $\mathbb{A}$ as a map $a \curvearrowright l_a$ of $\mathbb{A}$ into $\mathbb{R}_+$ such that Kraft's inequality holds in idealized form. The $l_a$'s are thought of as code-word lengths and the idealization lies in accepting arbitrary real values for the $l_a$'s and also in insisting that equality holds in (**??**). This last requirement aims in itself at efficiency by not allowing the $l_a$'s to be unnecessarily large. The use of code lengths of non-inter values may be justified by the use of block coding, where blocks of input letters are encoded instead of individual input letters. For block codes one should consider the average codelength per input letter and this will obviously lead to non-integer values.

From now on we will identify a code with a length function $l : \mathbb{A} \rightarrow \mathbb{R}_+$ such that Kraft's inequality is satisfied. How the actual code $\kappa : \mathbb{A} \rightarrow \{0, 1\}^*$ looks like will not be discussed

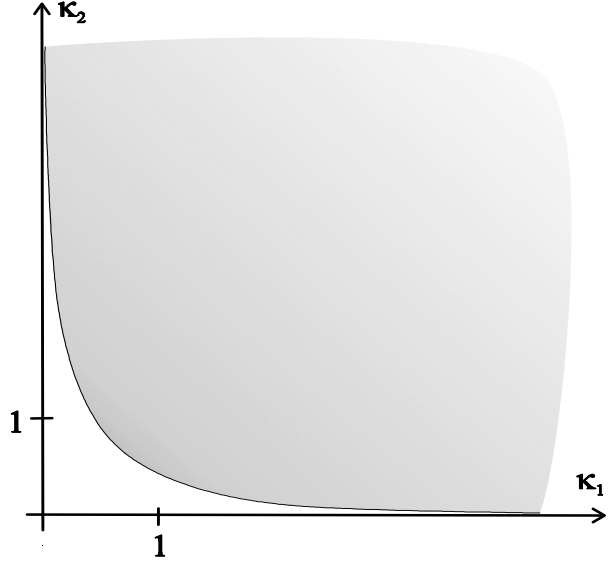

FIG. 1: Kraft's inequality is satisfied in the shaded area. The full curve indicates the compact codes.

in any more detail. Note that with these definitions the set of codes is convex.

Let $l : \mathbb{A} \to \mathbb{R}_+$ denote a code. Put

$$Z = \sum_{a \in \mathbb{A}} 2^{-l_a}. \tag{2}$$

This quantity is closely related to the partition function used in statistical mechanics. Now $\log Z \leq 0$ and $l + \log Z$ is a code for which Kraft's inequality is fulfilled with equality. Thus $l + \log Z$ can be considered as a compressed version of the code $l$ in the sense the all code words have become shorter. Note that Kraft's inequality for the code $l$ holds with equality, if and only if the inequality $l'_a \leq l_a$ for all $a \in \mathbb{A}$ implies that $l' = l$. Codes where Kraft's inequality holds with equality are called *compact codes*. Apparently, there is a one-to-one relationship between compact codes and probability distributions. It is given by the formulas

$$l_a = -\log p_a \; ; \; p_a = 2^{-l_a}. \tag{3}$$

When these formulas hold, we say that the code is *adapted to $P$* or vice versa.

In order to analyze compact codes the following definition is useful.

**Definition 1** *Let $P$ and $Q$ be probability measures. For $q \in \,]0;1[$ the the Rényi divergence from $P$ to $Q$ of order $q$ is defined by the equation*

$$D_q\left(P\|Q\right) = \frac{1}{1-q} \log \left( \int \left(\frac{dP}{dQ}\right)^q dQ \right). \tag{4}$$

**Proposition 2** *Assume that $\kappa_1$ and $\kappa_2$ are compact codes and $P_1$ and $P_2$ are the corresponding probability measures. For $q \in [0; 1]$*

$$(1-q)\,\kappa_1 + q\kappa_2 - qD_{1-q}\,(P_1 \| P_2) \tag{5}$$

*is a compact code.*

**Proof.** We have to check that Kraft's inequality holds with equality. We have

$$\sum_{a \in \mathbb{A}} \exp\left(-\left(1-q\right)\kappa_1\left(a\right) - q\kappa_2\left(a\right)\right) = \sum_{a \in \mathbb{A}} P_1^{1-q}\left(a\right)P_2^q\left(a\right) \tag{6}$$

$$= \sum_{a \in \mathbb{A}} \left(\frac{P_1\left(a\right)}{P_2\left(a\right)}\right)^{1-q} P_2\left(a\right).$$

∎

We see that for $q \in ]0; 1[$ the Rényi divergence measures how much a probabilistic mixture of two codes can be compressed. This shall be our operatioal definition of Rényi divergence. By use of Formula 4 the Rényi divergence can also be defined for values of $q$ greater than 1. For $q$ tending to 1 the Rényi divergence tends to

$$\sum_{a \in \mathbb{A}} p_a \log\left(\frac{p_a}{q_a}\right), \tag{7}$$

which is normally called the *information divergence from $P$ to $Q$* and is denoted $D\left(P \| Q\right)$. In the literature information divergence is sometimes termed *relative entropy* and denoted $H\left(Q, P\right)$ or $S\left(Q, P\right)$. This has created a lot of confusion because information divergence and entropy have opposite convexity properties. Information divergence is also some times called *Kullback-Leibler discrimination information* and denoted $I\left(Q, P\right)$. Unfortunately this notation is easily confused with the notation of *mutual information* and therefore we will follow I. Csiszár and use the notation $D\left(P \| Q\right)$.

For $q$ tending to 0 the Renyi entropy converges to the logarithm of the number of points where both $P$ and $Q$ have positive point probability and this is by definition the Rényi entropy of order 0. Thus our operational definition of Rényi divergence extends to $q \in [0; 1]$.

## III.   PROPERTIES OF RÉNYI DIVERGENCE

For $q \in ]0; 1[$ the Rényi divergence $D_q\left(P \| Q\right)$ is joint convex in $P$ and $Q$. This can be checked directly by calculations but it is much more instructive to realize that the convexity

follows from Proposition 2 and the fact that the set of codes satisfying Kraft's inequality is convex. Convexity also holds for $q = 1$ by continuity, but in general $D_q(P\|Q)$ is not convex in $P$ and $Q$ for $q > 1$.

Convexity of the set of codes also implies that $D_q(P\|Q)$ is increasing in $q$ when $P$ and $Q$ are fixed. Actually $D_q(P\|Q)$ is increasing in $q$ on the whole set $]0;\infty[$.

Let $P_1$ and $Q_1$ be two probability measures on the same set and let $P_2$ and $Q_2$ be probability measures on another set. The operational characterization implies that

$$D_q(P_1 \times P_2 \| Q_1 \times Q_2) = D_q(P_1\|Q_1) + D_q(P_2\|Q_2).\qquad(8)$$

We say that Rényi divergence is *additive* or *extensive*. This property also holds for all $q \in ]0;\infty[$.

## IV. SHANNON ENTROPY

Consider a source generating symbols over an alphabet $\mathbb{A}$ according to a known probability distribution $P = (p_a)_{a \in \mathbb{A}}$. We are now ready to define the *Shannon entropy* of $P$:

$$H(P) = \min_\kappa \sum_{a \in \mathbb{A}} p_a l_a,\qquad(9)$$

it being understood that the minimum is over all id-codes $\kappa$ (with the $l_a$'s denoting the idealized code-word lengths). Thus, *entropy is minimal average code-word length* understood in an idealized sense. A key result is the analytical identification of entropy:

**Theorem 3 (First main theorem of information theory)** *Denoting logarithms to the base 2 by* $\log$*, the entropy of* $P$ *given by (9) can be expressed analytically as follows:*

$$H(P) = -\sum_{x \in \mathbb{A}} p_x \log p_x.\qquad(10)$$

The relation of entropy to coding was emphasized by introducing the concept of idealized codes. By Theorem 3, the id-code adapted to $P$ is the optimal id-code of a source governed by $P$.

The idealization in Theorem 3 is a great convenience and no serious restriction. To emphasize this, let us insist, instead, to use codes with integer lengths. Then we can choose code-lengths $l_x$ close to $-\log p_x$ and ensure in this way that $H(P) \leq \sum p_x l_x < H(P) + 1$.

Moreover, if we consider a source generating sequences of letters independently according to the distribution $P$, then the minimum average code-word length per letter when we consider longer and longer sequences of letters converges to $H(P)$.

Let $U$ denote the uniform distribution on $\mathbb{A}$. Then the Shannon entropy and the information divergence are related by

$$H(P) = H(U) - D(P\|U).\tag{11}$$

In this sense the Shannon entropy measures deviation from being uniform.

Let $X$ and $Y$ be random variables with joint distribution $P$ and marginal distributions $P_X$ and $P_Y$. Then the mutual information between $X$ and $Y$ is given

$$I(X;Y) = D(P\|P_X \times P_Y).\tag{12}$$

Thus the mutual information measures how much the joint distribution deviates from being a product measure, or equivalently how much $X$ and $Y$ deviates from being independent. A significant observation is that we may characterize entropy as *self-information* since, for $Y = X$, we have

$$H(X) = I(X;X).\tag{13}$$

Thus entropy can be identified with self-information.

## V. RÉNYI ENTROPY

Entropy, or rather *entropy differences*, may be defined directly from divergence using the guiding equation

$$D_q(P\|U) = H_q(U) - H_q(P),\tag{14}$$

with $U$ the uniform distribution over $\mathbb{A}$. This only works with a finite alphabet $\mathbb{A}$. One may obtain absolute entropy if one adds the assumption that the entropy of a uniform distribution for any sensible notion of entropy must be the *Hartley entropy*, the logarithm of the size of the alphabet. Doing that, one finds that (14) leads to the quantity

$$H_q(P) = \frac{1}{1-q}\ln\sum_{a\in\mathbb{A}} p_a^q\tag{15}$$

which is the *Rényi entropy of order q*.

By using the properties of Rényi divergence we see that $H_q(P)$ is concave in $P$ for $q \in [0;1]$, but concavity breaks down for $q > 1$. We also see that $H_q$ decreasing in $q$ and is additive/extensive for all $q$. Concavity of entropy of order $q > 1$ can be reestablished by replacing the Rényi entropy by the Tsallis entropy but then additivity/extensivity is lost.

If $X$ and $Y$ are random variables with joint distribution $P$ and marginal distributions $P_X$ and $P_Y$ it is natural to define the mutual information of order $q$ by

$$I_q(X;Y) = D_q(P\|P_X \times P_Y).\tag{16}$$

Thus the Rényi self-information $X$ should be defined as $I_q(X,X)$. It is important to remark that Rényi entropy and Rényi self-information of order $q$ do not equal but are related by the equation

$$H_q(X) = I_{2-q}(X,X).\tag{17}$$

Only for $q = 1$ entropy and self-information are equal. Thus, leaving the classical Shannon case, it appears that entropy "splits up" in $H_q$ and $H_{2-q}$. The "duality" between order $q$ and order $2-q$ appears in many formulas and is related to the so-called escort probabilities.

For $q \in \, ]0;1[$ we have an operational definition of Rényi divergence, and these can easily be translated into operational definitions of Rényi entropies of order $q \in \, ]0;2[ \setminus \{1\}$ using Equation 14 or 17.

## VI. CONCLUDING REMARKS

Kraft's inequality leads directly to an operational definition of Rényi divergence of order $q \in [0;1]$ and to Rényi entropy of order $[0;2]$. Many of the fundamental properties of Rényi divergence and Rényi entropy can be derived directly from the operational definition. The relation between the operational defintions given here and cut-off rates will be discussed in a subsequent paper.